\documentclass[twocolumn,showpacs,preprintnumbers,amsmath,amssymb,superscriptaddress]{revtex4}

\usepackage{amssymb}
\usepackage{amsfonts}
\usepackage{amsmath}
\usepackage{graphicx}
\usepackage{textcomp}
\usepackage{mathptmx}
\usepackage{dcolumn}
\usepackage{bm}
\usepackage{times}

\begin{document}

\title{Observation of high coherence in Josephson junction qubits \\
measured in a three-dimensional circuit QED architecture}

\author{Hanhee Paik}
\affiliation{Department of Physics and Applied Physics, Yale University, New Haven, Connecticut 06520, USA}
\author{D. I. Schuster}
\affiliation{Department of Physics and Applied Physics, Yale University, New Haven, Connecticut 06520, USA}
\affiliation{Department of Physics and James Franck Institute, University of Chicago, Chicago, Illinois 60637, USA}
\author{Lev S. Bishop}
\affiliation{Department of Physics and Applied Physics, Yale University, New Haven, Connecticut 06520, USA}
\affiliation{Joint Quantum Institute and Condensed Matter Theory Center, Department of Physics, University of Maryland, College Park, Maryland 20742, USA}
\author{\\G. Kirchmair}
\author{G. Catelani}
\author{A. P. Sears}
\affiliation{Department of Physics and Applied Physics, Yale University, New Haven, Connecticut 06520, USA}
\author{B. R. Johnson}
\affiliation{Department of Physics and Applied Physics, Yale University, New Haven, Connecticut 06520, USA}
\affiliation{Raytheon BBN Technologies, Cambridge, MA 02138, USA}
\author{M. J. Reagor}
\author{\\L. Frunzio}
\author{L. I. Glazman}
\author{S. M. Girvin}
\author{M. H. Devoret}
\author{R. J. Schoelkopf}
\affiliation{Department of Physics and Applied Physics, Yale University, New Haven, Connecticut 06520, USA}
\date{\today}

\begin{abstract}
Superconducting quantum circuits based on Josephson junctions have made rapid progress in demonstrating quantum behavior and scalability. However, the future prospects ultimately depend upon the intrinsic coherence of Josephson junctions, and whether superconducting qubits can be adequately isolated from their environment.  We introduce a new architecture for superconducting quantum circuits employing a three dimensional resonator that suppresses qubit decoherence while maintaining sufficient coupling to the control signal.  With the new architecture, we demonstrate that Josephson junction qubits are highly coherent, with $T_2 \sim 10~\mu$s to $20~\mu$s without the use of spin echo, and highly stable, showing no evidence for $1/f$ critical current noise.  These results suggest that the overall quality of Josephson junctions in these qubits will allow error rates of a few $10^{-4}$, approaching the error correction threshold.
\end{abstract}


\maketitle

Superconducting circuits are a promising technology for quantum information processing with solid-state devices. Several different types of qubits \cite{Clarke08a, Schoelkopf08a} have been developed, which all rely on the nonlinearity of one or more Josephson junctions. Ideally, the Josephson junctions should be dissipationless and highly stable to avoid decoherence, while providing the crucial anharmonicity that allows individual energy levels to be separately addressed. In the past decade, the coherence time of superconducting qubits has increased from initially only a few nanoseconds to typically about a microsecond today. This has permitted experiments where two or three qubits are controlled, entangled \cite{Chow10b, Ansmann10a, DiCarlo10a, Neeley10a}, and used to demonstrate simple algorithms \cite{DiCarlo09a}.  However, scaling more than three qubits with acceptable level of fidelity and coherence will require higher coherence times than the current state-of-art.  Two major outstanding questions are whether superconducting qubit coherence can improve further and whether there are fundamental limits on coherence imposed by the Josephson junctions.

The coherence can either be limited by possible imperfections in the Josephson junctions, or by unintended interactions with the environment.  Even if the junctions were perfectly coherent, achieving a long coherence time also requires understanding and controlling the Hamiltonian such that the terms coupling the qubit to the outside world can be made small. For example, in the hydrogen atom a coupling of only 40 parts per billion (ppb) to the electromagnetic continuum gives rise to spontaneous emission and a quality factor, Q, of about 25 million. Building a scalable quantum computer using superconducting qubits therefore requires engineering a Hamiltonian where the undesirable couplings that lead to decoherence are kept at the part per million (ppm) to ppb level. Can a manmade, macroscopic quantum system based on Josephson junctions have well-defined quantum states that approach this level of coherence?

Here we present results on a new implementation of a superconducting qubit where we carefully control the coupling to the environment, obtaining an increase in coherence by over an order of magnitude. We observe reproducible qubit lifetimes for relaxation ($T_1$) up to 60~$\mu\rm{s}$, and lifetimes of coherent superpositions ($T_2$) of 10 - 20 $\mu\rm{s}$, corresponding to quality factors for both dissipation ($Q_1 = \omega_{01} T_1 \sim 2 \times 10^6$, where $\omega_{01}$ is the transition frequency of the qubit) and decoherence ($Q_2 = \omega_{01} T_2 \sim 7 \times 10^5$) of about one million. The high quality factors observed imply that the instability in the parameters of our system, the intrinsic dissipation in the Josephson junction, and the size of the undesired couplings are all in fact smaller than one ppm. Together with the fast gate time ($t_{gate} ~ \sim$ 10 ns) previously demonstrated in superconducting qubits \cite{Chow10a}, we estimate the error rate will be approximately $t_{gate}/T_2 \sim 5 \times 10^{-4}$.  These results suggest that existing Josephson junction technology should allow superconducting circuits to achieve coherence levels compatible with scalable quantum computing in the solid state.

Our experiment employs a particularly simple transmon qubit \cite{Koch07a,Schreier07a}, consisting of just two superconducting electrodes connected with a single small aluminum Josephson junction, that requires no bias circuitry and has minimal sensitivity to $1/f$ noises in charge or flux, coupled to a microwave resonant cavity that can act as an entanglement bus and readout circuit. Neglecting the interactions with its environment, the transmon is described by the simple Hamiltonian \cite{Devoret04a, Koch07a} $\hat{H} = 4E_C(\hat{n}-n_0)^2 - E_J \cos\hat{\phi}$ where $\hat{n}$ and $\hat{\phi}$ are the normalized operators for the pair charge and phase (obeying $[\hat{\phi},\hat{n}]=i$), $E_J = \hbar I_c/2e$ and $E_C = e^2/2C_\Sigma$ are the Josephson and Coulomb energies, $e$ is the electron's charge, $I_c$ is the junction critical current, $C_\Sigma$ is the total capacitance between the electrodes, and $n_0$ is the offset charge.

\begin{figure}
\includegraphics[width=3.5in]{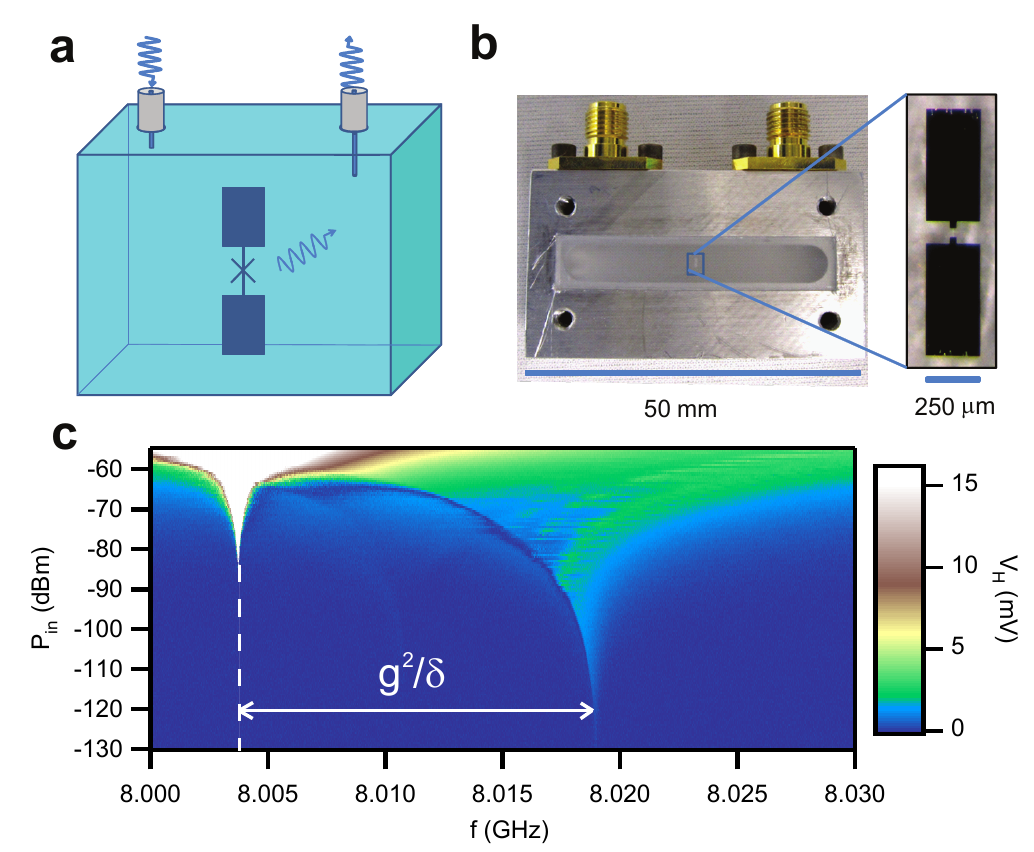}
\caption{\label{fig1ver7} Qubit coupled to a 3D cavity (a) Schematic of a transmon qubit inside a 3D cavity.  The qubit is coupled to the cavity through a broadband dipole antenna that is used to receive and emit photons.  (b) Photograph of a half of the 3D aluminum waveguide cavity.  An aluminum transmon qubit with the dipole antenna is fabricated on a c-plane sapphire substrate and is mounted at the center of the cavity.  (Inset) Optical microscope image of a single-junction transmon qubit. The dipole antenna is 1 mm long. (c)  Transmission of a 3D cavity (cavity D) coupled to a transmon ($J1$) measured as a function of power and frequency.  The cavity response above -80 dBm occurs at the bare cavity frequency $f_c$ = 8.003 GHz. At lower powers, the cavity frequency shifts by $g^2/\delta$. }\label{fig1}
\end{figure}

\begin{table*}
\begin{tabular}{c|cccccccccc}
$\underset{(cavity)}{qubit}$ & $\underset{(\mathrm{GHz})}{f_{01}}$ & $\underset{(\mathrm{GHz})}{E_J} $ &
$\underset{(\mathrm{GHz})}{E_{C}} $  & $\underset{(\mathrm{MHz})}{g/2\pi} $ & $\underset{(\mathrm{MHz})}{g^2/2\pi\delta}$ &
$\underset{(\mathrm{GHz})}{f_c}$  & $\underset{(x10^{3})}{Q_{c}}$  & $\underset{(\mu \mathrm{s})}{T_1}$ & $\underset{(\mu \mathrm{s})}{T_2}$ &  $\underset{(\mu \mathrm{s})}{T_{\mathrm{echo}}}$  \tabularnewline
\hline
J1 (D)     & 6.808   & 21.1   & 0.301 & 138 &  15.9 & 8.0035 & 340 & 60  & 18 & 25  \tabularnewline
J1a (D)     & 6.769   & 21.0   & 0.301 & 140 &  15.8 & 8.00375& 340 & 50  & 20 & 24  \tabularnewline
J2 (C)     & 7.772   & 28.6   & 0.292 & 152 &  99.8  & 8.0020 & 360 & 25  & 15  & 21 \tabularnewline
J3 (B)     & 7.058   & 22.5  & 0.304 & 141 &  21.5 & 7.9835 & 320 & 42 & 12  & 12  \tabularnewline
S (D)      & 7.625  &  34.4  & 0.227 & 136 & 48.2 & 8.01065 & 340 & 35  & 7.3  & 11   \tabularnewline
Sa (A)     & 7.43   &  32.5   &  0.228 & 123 & 24.1 & 8.06169 & 100 & 20  & 6  & 8 \tabularnewline
\end{tabular}
\caption{\label{table1} Parameters of four transmon qubits (labeled as $J$'s for single-junction qubits and $S$'s for SQUID) measured in four different 3D cavities (labeled as A, B, C and D respectively).  The data of $J1a$ and $Sa$ are data on qubits J$1$ and $S$ following cycling to room temperature.  Here, $f_{01} = \omega_{01}/2\pi$ is the dressed qubit transition frequency between ground state $|0\rangle$ and the first excited state $|1\rangle$, $g/2\pi$ is the coupling strength, $g^2/2\pi\delta$ is the cavity frequency shift from the bare cavity frequency due to the qubit, $f_c = \omega_c/2\pi$ is the bare cavity frequency, $Q_c$ is the quality factor of the shifted cavity resonance at single-photon power, $T_1$ is the relaxation time from $|1\rangle$ to $|0\rangle$, $T_2$ is the coherence time measured by Ramsey experiment, and $T_{\mathrm{echo}}$ is the coherence time measured by a spin-echo experiment.}
\end{table*}
	
The experiments are performed using a circuit QED architecture \cite{Wallraff04a, Blais04a}, a circuit implementation of a cavity QED \cite{Haroche06}, to isolate, couple, and measure the qubit. A novel aspect is the use of a three-dimensional waveguide cavity machined from superconducting aluminum (alloy 6061 T6), as shown in Fig. 1a-1b. This type of cavity offers several advantages over the planar transmission-line cavities used in previous circuit QED experiments. First, the cavity has a larger mode volume (approximately 3 cm$^3$ or one tenth of a cubic wavelength, compared to the $10^{-6}$ cubic wavelengths for a conventional transmission line resonator), and is much less sensitive to the surface dielectric losses that are suspected as the limiting source of dissipation in transmission line resonators to date \cite{Gao98a, OConnell98a}. Indeed, we have observed reproducible quality factors of these cavities \cite{Paik11a} of 2 to 5 million, corresponding to photon storage times in excess of 50 $\mu$s (not shown) in the quantum regime ($k_B T \ll \hbar\omega_{c}$ and $\langle n\rangle < 1$, where $\omega_c$ is the cavity frequency), without the power dependence \cite{Gao98a, OConnell98a} indicative of the presence of two-level systems. Second, the geometry presents the qubit with a well-controlled electromagnetic environment, limiting the possibility of relaxation through spontaneous emission into the multiple modes that may be possible with a complicated chip and its associated wiring \cite{Houck08a}. The qubit is placed in the center of the cavity, maximizing the coupling to the lowest frequency TE101 mode at $\omega_c/2\pi\ \sim$~8 GHz, which is used for readout and control. This location also nulls the coupling to the second mode (TE102 at approximately 10 GHz).

Despite the larger mode volume of the three-dimensional cavity, we are able to achieve the strong-coupling limit of cavity QED in this system, with vacuum Rabi frequencies, $g/2\pi$, greater than 100 MHz. As seen in Figure 1b, the electrodes of the qubit are significantly larger ($\sim 0.5$ mm) than in a conventional transmon qubit, so that the increased dipole moment of the qubit compensates for the reduced electric field that a single photon creates in the cavity. We note that due to the large dipole moment, the expected lifetime from spontaneous emission in free space would be only $\sim$ 100 ns, so that a high-Q cavity is required to maintain the qubit lifetime. The electrodes also form the shunting capacitance ($C_\Sigma\sim$ 70 fF) of the transmon, giving it the same anharmonicity and the same insensitivity to $1/f$ charge noise as in the conventional design. An advantage of this qubit design is that the large electrode size reduces the sensitivity of the qubit to surface dielectric losses, which may be responsible for the improved relaxation times. In this experiment, the qubits cannot be tuned into resonance with the cavity, so the vacuum Rabi coupling is not observed directly. The system is rather operated in the dispersive limit ($|\delta|=|\omega_c-\omega_{01}| \gg g$) \cite{Blais04a}. Here the qubit induces a state-dependent shift on the cavity, which is the basis of the readout mechanism. The dispersive shifts are typically several tens of MHz (see Table 1), and can approach 1,000 times the linewidths of qubit and cavity, so that all devices are well within the strong dispersive limit \cite{Schuster07a}. The transmission through the cavity as a function of microwave power, which demonstrates the ground-state shift of the cavity and the re-emergence of the bare cavity frequency at sufficiently high powers (see Ref.\cite{Reed10a, Bishop10a}) is shown in Figure 1c. Single-shot readout of the qubit (with fidelities greater than 70\%) is performed using the technique previously described \cite{Reed10a}.

\begin{figure}
\includegraphics[width=3in]{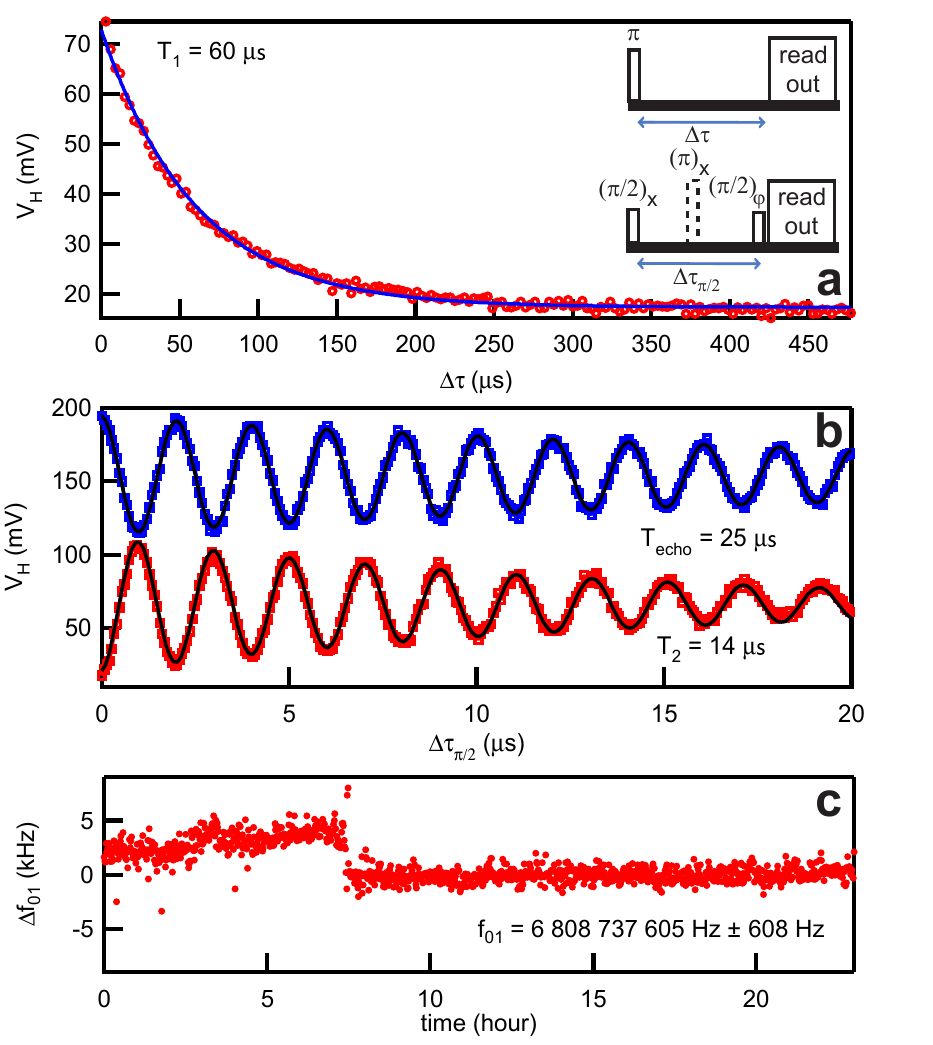}\caption{Time domain measurement of the qubit coherence (a) Relaxation from $|1\rangle$ of qubit $J1$.  $T_1$ is 60 $\mu s$ for this measurement.  (Inset) The pulse sequences used to measure relaxation (upper) and Ramsey experiments with and without echo (lower).  The pulse shown in dashed line is an echo signal applied at one half of the delay between two $\pi/2$ pulses.  (b) Ramsey fringes measured on resonance with (blue squares) and without (red squares) echo sequence.  The pulse width for the $\pi$ and $\pi/2$ pulses used in the experiments is 20 ns. An additional phase is added to the rotation axis of the second $\pi$/2 pulse for each delay to give the oscillatory feature to the Ramsey fringes. (c) Qubit frequency $f_{01}$ measured over 23 hours by repeated Ramsey measurements.}\label{fig2}
\end{figure}

The dramatically improved coherence properties of these qubits are confirmed via the standard time-domain measurements of the relaxation time ($T_1$) and Ramsey experiments ($T_2$) (see Figure \ref{fig2} and Table~1).  We employ the same techniques used in the previous conventional transmon experiment (see Ref.\cite{Houck08a, Chow10a, sup1}) performed in a cryogen-free dilution refrigerator at 10 mK.  The qubits have an anharmonicity $\alpha/2\pi = f_{12} - f_{01} \sim$ $-200$~MHz to $-300$~MHz which allows fast single-qubit operations in $\sim 10$~ns.  There are several surprising features in the time-domain data.  First, while $T_2$'s are typically in the range of 15 - 20 $\mu$s, they do not yet attain the limit twice $T_1$ which is reproducibly in the range 25 - 50~$\mu$s corresponding to $Q_1$= 1 - 2$\times 10^6$. This indicates that there is still significant dephasing. At the same time, both the Ramsey decay envelope and the echo coherence (which has an artificial phase rotation as a function of the delay added) can be fit well by an exponential decay, indicating that contrary to the previous predictions \cite{Harlingen04a, Eroms06a}, $1/f$ noise is not dominant \cite{sup1} in our experiment. This is consistent with the expectation that these simple qubits should avoid dephasing due to both $1/f$ flux noise (since there are no superconducting loops) and charge noise (since the total charge variation of the transition frequency \cite{Koch07a} for these transmon parameters is less than 1~kHz). The observed phase coherence factor, $Q_2 \sim 700,000$, is an order of magnitude larger than previous superconducting devices \cite{Vion04a, Schreier07a}. Since the transition frequency of the transmon qubit ($\omega_{01} \sim \sqrt{8E_J E_C})$ is set by a combination of the Josephson and charging energies, this allows us to place stringent limits on the amount of critical current noise in Josephson junctions, or dielectric fluctuations in the shunt capacitance. We find that any critical current noise must have a total variance of less than about one ppm ($\delta I/I_c < 10^{-6}$). In fact, the observed Ramsey and echo decays are more consistent with frequency-independent noise in the qubit transition frequency (or critical current) of about $\sqrt{S_{\omega/\omega_{01}}} \sim 10\ {\rm ppb} /\sqrt{{\rm Hz}}$ or $\sqrt{S_{I/I_c}} \sim 20\ {\rm ppb} /\sqrt{{\rm Hz}}$. The improvement by approximately a factor of two with echo indicates either a characteristic correlation time in this noise of about 10~$\mu$s, or an additional low-frequency noise (but not $1/f$ in character) component with a variance of about 1 ppm.

The high stability of the qubit transition frequency is exhibited in Fig. \ref{fig2}c, which shows the deviations observed in the Ramsey detuning compared to the microwave generator over one day. The parameters in the Hamiltonian of this artificial quantum system are seen to be stable for long periods to within $\sim 600$~Hz or $\sim 80$~parts per billion over many hours. Discrete jumps of a few kHz (or about 1 ppm) are occasionally observed, which are consistent in size with that expected by single atomic rearrangements in the tunnel junction barrier. On subsequent thermal cycling of two devices, a slow telegraph switching behavior (with $\delta f \sim 5 - 50$~kHz) was also observed. These observations confirm the ability of this type of experiment to reveal tiny variations in junction parameters which would be undetectable in previous superconducting qubit experiments.

\begin{figure}
\includegraphics[width=3.4in]{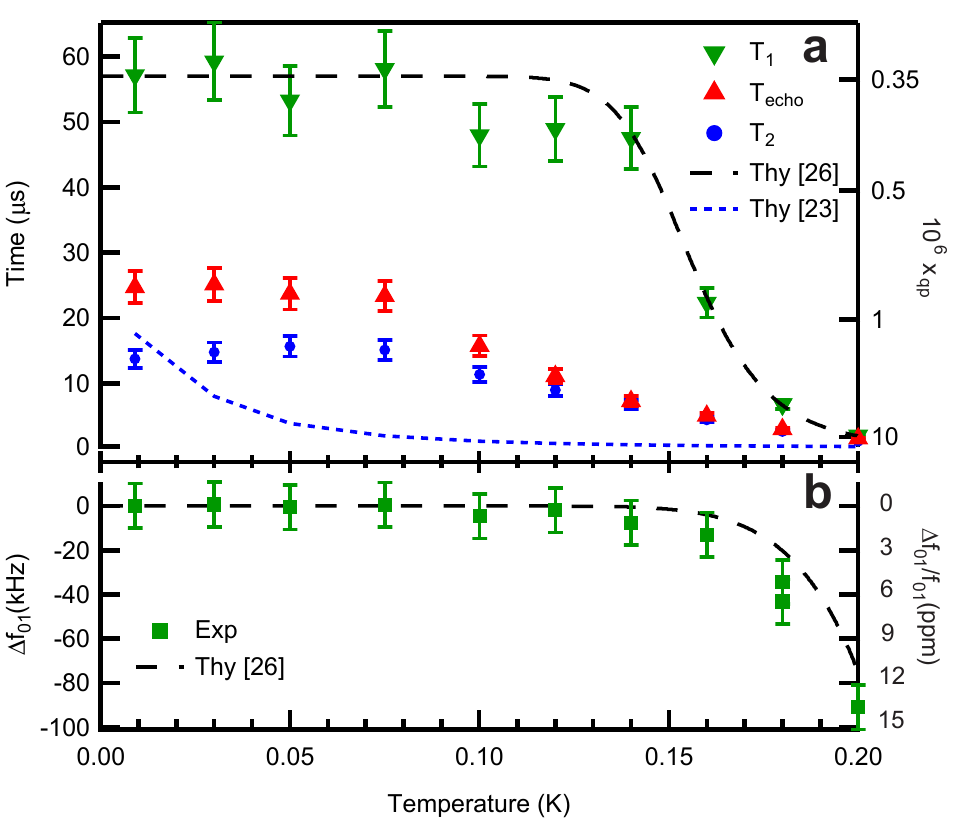}
\caption{Temperature dependence of qubit properties (a) Measurement of $T_1$, $T_{\mathrm{echo}}$ and $T_2$ (b) Shift of the transition frequency $f_{01}$. Black dashed curves in (a) and (b) are theoretical $T_1$ and $f_{01}$ calculated from the model in Ref.\cite{Catelani11a} plus a temperature-independent relaxation rate with the same fitting parameter of $\Delta$ = 194 $\mu$eV for both $T_1$ and $f_{01}$ (See supplement for details).  Blue dashed curve is the theoretical $T_2$ from $1/f$ critical current fluctuations predicted by Ref.\cite{Harlingen04a}}\label{fig3}
\end{figure}

It is still not clear what mechanisms limit the coherence in these new qubits, but some information can be obtained by measuring the dependence of coherence times on sample temperature, as shown in Figure \ref{fig3}. The most striking effect is the rapid decrease in the relaxation time, $T_1$, once the temperature exceeds about 130~mK, which is in good quantitative agreement with a recent theory on the effects of quasiparticles \cite{Catelani11a, sup1}, using only a single fit parameter which is the gap of the superconductor ($\Delta$ = 194 $\mu$eV). The saturation of the relaxation times below this temperature means that either the qubits are limited by the presence of out-of-equilibrium quasiparticles, or other mechanisms such as spontaneous emission (the Purcell effect) or dielectric losses in the sapphire. The observed times place a stringent upper bound on the density of these quasiparticles, which is less than one quasiparticle per cubic micron, or a normalized quasiparticle density $x_{qp} = n_{qp}/n_{pairs} < 5 \times 10^{-7}$ (where $n_{qp}$ is the density of quasiparticles and $n_{pairs}$ is the density of Cooper pairs). This is an order of magnitude lower density than reported in recent measurements on phase qubits \cite{Lenander11a} or transmission-line resonators \cite{Visser11a}.  We also observe a frequency shift of the qubit transition due to the induced thermal quasiparticles (Fig 3b), again in quantitative agreement with the predictions of Reference~\cite{Catelani11a}. The coherence times $T_2$ and $T_{\mathrm{echo}}$ are observed to decrease slowly with temperature, inconsistent with either the quadratic \cite{Harlingen04a} or linear \cite{Eroms06a} temperature scalings reported previously for critical current noise. Another mechanism for relaxation is spontaneous emission of the qubit via the cavity (the Purcell effect), which is consistent with the general trend that qubits with smaller detunings from the cavity tend to have shorter lifetimes. Our estimates of the multi-mode Purcell effect \cite{Houck08a} for these cavities predict only a small contribution to $T_1$ for most devices measured so far, but the data do not yet allow a detailed test of the Purcell model. Further experiments which vary qubit and cavity parameters will be required to identify and hopefully further reduce decoherence. Nonetheless, the current results already indicate that the intrinsic quality factor of Josephson junctions is greater than one million.

We have presented a new implementation of superconducting qubits coupled to a 3D microwave cavity with an order of magnitude improved coherence.  While there have been recent observations of individual devices with comparable $T_1$ \cite{Kim11a} or $T_{\rm echo}$ times \cite{Bylander11a}, we have reproducibly obtained both relaxation ($T_1$) and coherence ($T_2$, without echo) times in excess of 10 $\mu$s.  Our qubits demonstrate remarkable long-term stability, indicating that critical current fluctuations are much smaller than previously predicted. The dissipation and the frequency shift as a function of temperature confirm recent theoretical predictions on the effects of thermally-induced quasiparticles.  The lifetimes at low temperatures imply that the intrinsic quality factor of Josephson junctions can be greater than one million and enable us to place significantly more stringent limits on any possible background density of non-equilibrium quasiparticles.

Our new architecture provides a particularly simple electromagnetic environment for the qubits, and thereby reduces the sources of decoherence.   The good stability and long coherence times obtained from the new architecture enable us to detect couplings of the qubit to any low-energy degrees of freedom, at the level of a fraction of only a ppm.  The improved coherence was achieved without reducing either the anharmonicity or the coupling strength between the qubit and cavity which should still permit fast one and two-qubit operations.  Scaling this architecture to multiple qubits is not harder than for conventional superconducting circuits.  For example, more qubits can be added inside a 3D cavity such that they couple to each other.  These results are therefore encouraging for all experiments with superconducting quantum circuits.  The evolution of this architecture could allow future devices to approach the error levels required to achieve the quantum error correction threshold and make it possible to realize larger entangled states and more complex algorithms with superconducting quantum systems.

We thank Eran Ginossar, Matt Reed, Leo DiCarlo, Luyan Sun and Shyam Shankar for valuable discussions. L. F. acknowledges partial support from CNR-Istituto di Cibernetica. This research was funded by the Office of the Director of National Intelligence (ODNI), Intelligence Advanced Research Projects Activity (IARPA), through the Army Research Office.  All statements of fact, opinion or conclusions contained herein are those of the authors and should not be construed as representing the official views or policies of IARPA, the ODNI, or the U.S. Government.


\begin{thebibliography}{31}

\bibitem{Clarke08a} J. Clarke and F. K. Wilhelm, \textit{Nature} \textbf{453}, 1031-1042 (2008).

\bibitem{Schoelkopf08a} R. J. Schoelkopf and S. M. Girvin, \textit{Nature} \textbf{451}, 664-669 (2008).

\bibitem{Chow10b} J. M. Chow \textit{et al.}, Phys.\ Rev.\ A \textbf{81}, 062325 (2010).

\bibitem{Ansmann10a} M. Ansmann \textit{et al.}, Nature \textbf{461}, 504-506 (2009).

\bibitem{DiCarlo10a} L. DiCarlo \emph{et al.}, Nature \textbf{467}, 574-578 (2010).

\bibitem{Neeley10a} M. Neeley \emph{et al.}, Nature \textbf{467}, 570-573 (2010).

\bibitem{DiCarlo09a} L. DiCarlo \emph{et al.}, Nature \textbf{460}, 240-244 (2009).

\bibitem{Chow10a} J. Chow \textit{et al.}, Phys.\ Rev.\ A \textbf{82}, 040305(R) (2010).

\bibitem{Koch07a} J. Koch \textit{et al.}, Phys.\ Rev.\ A \textbf{76}, 042319 (2007).

\bibitem{Schreier07a} J. A. Schreier \textit{et al.}, Phys.\ Rev.\ B \textbf{77}, 180502(R) (2008).

\bibitem{Devoret04a} M. H. Devoret and J. M. Martinis, \textit{Quant. Info.  Proc.} \textbf{3}, 163-203 (2004).

\bibitem{Wallraff04a} A. Wallraff \textit{et al.}, Nature \textbf{431}, 162-167 (2004).

\bibitem{Blais04a} A. Blais, R.-S. Huang, A. Wallraff, S. M. Girvin, R. J. Schoelkopf,  Phys.\ Rev.\ A \textbf{69}, 062320 (2004).

\bibitem{Haroche06} S. Haroche and J.-M. Raimond, \textit{Exploring the Quantum: Atoms, Cavities, and Photons} (Oxford University Press, 2006).

\bibitem{Gao98a} J. Gao \textit{et al.}, Appl.\ Phys.\ Lett. \textbf{92}, 152505 (2008).

\bibitem{OConnell98a} A. D. O'Connell \textit{et al.}, Appl.\ Phys.\ Lett. \textbf{92}, 112903 (2008).

\bibitem{Paik11a} Paik \textit{et al.}, manuscript in preparation (2011).

\bibitem{Houck08a} A. A. Houck \textit{et al.}, Phys.\ Rev.\ Lett. \textbf{101}, 080502 (2008).

\bibitem{Schuster07a} D. I. Schuster \textit{et al.}, Nature \textbf{445}, 515-518 (2007).

\bibitem{Reed10a} M. D. Reed \textit{et al.}, Phys.\ Rev.\ Lett. \textbf{105}, 173601 (2010).

\bibitem{Bishop10a} L. S. Bishop, E. Ginossar, S. M. Girvin, Phys.\ Rev.\ Lett. \textbf{105}, 100505 (2010).

\bibitem{sup1} See Supplemental Material.

\bibitem{Harlingen04a} D. J. Van Harlingen \textit{et al.}, Phys.\ Rev.\ B  \textbf{70},
064517 (2004).

\bibitem{Eroms06a} J. Eroms \textit{et al.}, Appl.\ Phys.\ Lett. \textbf{89}, 122516 (2006).

\bibitem{Vion04a} D. Vion \textit{et al.}, Science \textbf{296}, 886-889 (2002).

\bibitem{Catelani11a} G. Catelani \textit{et al.}, Phys.\ Rev.\ Lett. \textbf{106}, 077002 (2011).

\bibitem{Visser11a} P. J. de Visser \textit{et al.}, Phys.\ Rev.\ Lett. \textbf{106}, 167004 (2011).

\bibitem{Lenander11a} M. Lenander \textit{et al.}, Phys.\ Rev.\ B \textbf{84}, 024501 (2011).

\bibitem{Kim11a} Z. Kim, \textit{et al.} Phys.\ Rev.\ Lett. \textbf{106}, 120501 (2011).

\bibitem{Bylander11a} J. Bylander \textit{et al.} Nature Phys. \textbf{7}, 565 (2011).

\end{thebibliography}
\end{document}